\documentclass[english,prb,aps,floats,showpacs,floatfix]{revtex4}
\usepackage[T1]{fontenc}
\usepackage[latin1]{inputenc}
\usepackage{amsmath}
\usepackage{graphicx}
\usepackage{amssymb}

\makeatletter

\providecommand{\tabularnewline}{\\}

\usepackage{graphicx}
\usepackage{psfig}

\usepackage{babel}
\makeatother
\begin{document}

\title{Photoinduced absorption in disubstituted polyacetylenes: Comparison
of theory with experiments }

\author{Priya Sony and Alok Shukla}

\affiliation{Physics Department, Indian Institute of Technology, Powai, Mumbai
400076 INDIA}

\begin{abstract}
In a recently performed experiment Korovyanko $et$ $al$ {[}Phys.
Rev. B \textbf{67}, 035114 (2003){]} have measured the photo-induced
absorption (PA) spectrum of phenyl-disubstituted polyacetylenes (
PDPA) from 1$B_{u}$ and 2$A_{g}$ excited states. In 1$B_{u}$ PA
spectrum they identified two main features namely PA1 and PA2. While
in the 2$A_{g}$ spectrum they identified only one feature called
PA$_{g}$. In this paper we present a theoretical study of 1$B_{u}$
and 2$A_{g}$ PA spectra of oligo-PDPA's using correlated-electron
Pariser-Parr-Pople (P-P-P) model and various configuration interaction
(CI) methodologies. We compare the calculated spectra with the experiments,
as well as with the calculated spectra of polyenes of the same conjugation
lengths. Calculated spectra are in good agreement with the experiments.
Based upon our calculations, we identify PA1 as the m$A_{g}$ state
and PA$_{g}$ as the $nB_{u}$ state of the polymer. Regarding the
PA2 feature, we present our speculations. Additionally, it is argued
that the nature of excited states contributing to the $2A_{g}$-PA
spectra of oligo-PDPA's is qualitatively different from those contributing
to the spectra of polyenes.
\end{abstract}


\maketitle

\section{Introduction}

\label{intro}

Recently discovered class of conjugated polymers called phenyl-disubstituted
polyacetylenes (PDPA's)---obtained by substituting the side H atoms
of \emph{trans}-polyacetylene by phenyl derivatives---have exhibited
very interesting optical properties.\cite{tada1,tada2,liess,fujii1,gontia,sun,hidayat,fujii2}
Despite their structural similarities to \emph{trans}-polyacetylene---which
is well-known to be nonphotoluminescent---PDPA's are known to exhibit
photoluminescence (PL) with large quantum efficiency.\cite{gontia}
We explained this apparently contradictory behavior, in our earlier
papers,\cite{shukla1,shukla2,shukla3} in terms of reduced electron-correlation
effects caused by the delocalization of exciton in the transverse
direction because of the presence of phenyl rings. It was argued in
those papers that because of the reduced correlation effects, the
first two-photon excited state $2A_{g}$ state is higher in energy
than the first one-photon excited state $1B_{u}$ state in PDPA's,
rendering these materials photoluminescent.\cite{shukla1,shukla2,shukla3}
This ordering of excited states is exactly opposite to that in \emph{trans}-polyacetylene,
where relatively stronger correlation effects shift the $2A_{g}$
state lower than the $1B_{u}$ state, making the material weakly emissive.\cite{dixit,sumit}
Furthermore, we argued that in PDPA's, the delocalization of excitons
has couple of more important consequences on their optical properties:
(a) reduced optical gaps as compared to \emph{trans}-polyacetylene,
and (b) significant presence of transverse polarization in the photons
emitted during the PL in PDPA's, a prediction, which since then, has
been verified in several experiments.\cite{fujii2,gontia2}

Recently, Korovyanko \emph{et al.}\cite{koro} \emph{}have probed
the excited states of PDPA's using the photoinduced absorption (PA)
spectroscopy, leading to measurement of excited states so far unexplored
in linear optics. They measured the PA spectrum of oligo-PDPA's both
from the $1B_{u}$ as well as $2A_{g}$ excited states. Using the
dipole selection rules for centro-symmetric polymers such as PDPA's,
it is obvious that the PA spectrum from the $1B_{u}$ state will have
peaks corresponding to the higher $A_{g}$-type states, while those
in the $2A_{g}$ spectrum will have features corresponding to higher
$B_{u}$-type states. In the PA spectrum from the $1B_{u}$ state,
Korovyanko et al. report observing two prominent features, referred
as PA1 and PA2 by them.\cite{koro} While in the PA spectrum from
the $2A_{g}$ state, they observed only one feature which they called
the PA$_{g}$ peak.\cite{koro} To understand the nature of the excited
states leading to these PA peaks, however, one needs to perform detailed
theoretical calculations of the PA spectra of oligo-PDPA's of various
sizes, and compare the theoretical results to the experimental ones.
Since, it is well-known that the electron-correlation effects play
very important roles in proper description of the excited states of
conjugated polymers,\cite{dixit,sumit} the calculations must properly
take them into account. Studying these excited states will also enlighten
us about the nonlinear optical properties of PDPA's because some of
these excited states will also be visible in their nonlinear spectra,
such as two-photon absorption (TPA), and third-harmonic generation
(THG). It is with this aim in mind that we have undertaken a systematic
theoretical study of the PA spectra of oligo-PDPA's of varying sizes.
In the present work, we have used the Pariser-Parr-Pople (P-P-P) model
Hamiltonian and employed a configuration-interaction (CI) method based
approach to perform the correlated calculations of the low-lying excited
states, and the PA spectra, of this material. We also compare the
states visible in the PA spectra of PDPA's with those in their TPA
and the THG spectra computed by us recently.\cite{shukla-tpa,shukla-thg} 

The remainder of this paper is organized as follows. In section \ref{method}
we briefly describe the theoretical methodology used to perform the
calculations in the present work. Next in section \ref{results} we
present and discuss the calculated PA spectra of oligo-PDPA's. Finally,
in section \ref{conclusion} we summarize our conclusions.

\section{Theory}

\label{method}

\begin{figure}
\begin{center}\includegraphics[%
  width=4cm]{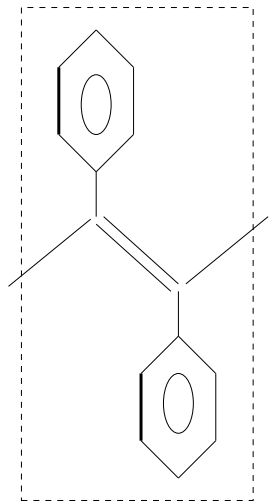}\end{center}

\caption{The unit cell of PDPA. The phenyl rings are rotated with respect
to the $y$-axis, which is transverse to the axis of the polyene backbone
($x$-axis)}

\label{fig-pdpa}
\end{figure}

The unit cell of PDPA oligomers considered in this work is presented
in Fig. \ref{fig-pdpa}.%
 To the best of our knowledge, the geometry of PDPA's in the ground
state, is still not known. However, it is intuitively clear that the
steric hindrance would cause a rotation of the substituent phenyl
rings so that they would no longer be coplanar with the backbone of
the polymer. The extent of this rotation is also unknown; however,
it is clear that the angle of rotation will be less than 90 degrees
because that would effectively disconnect the phenyl rings from the
backbone. In our previous works\cite{shukla1,shukla2,shukla3}, we
argued that the steric hindrance effects can be taken into account
by assuming that the phenyl rings of the unit cell are rotated with
respect to the $y$-axis by 30 degrees in such a manner that the oligomers
still have inversion symmetry. It is obvious that along the direction
of conjugation ($x$-direction), PDPA is structurally similar to \emph{trans-}polyacetylene,
with alternating single and double bonds. In the following, we will
adopt the notation PDPA-$n$ to denote a PDPA oligomer containing
$n$ unit cells of the type depicted in Fig. \ref{fig-pdpa}.

The point group symmetry of \emph{trans}-polyacetylene and polyenes
is $C_{2h}$ so that the one-photon states belong to the irreducible
representation (irrep) $B_{u}$, while the ground state and the two-photon
excited states belong to the irrep $A_{g}$. Because of the phenyl
group rotation mentioned above, the point group symmetry of PDPA's
and its oligomers is $C_{i}$ so that its ground state and the two-photon
excited states belong to the irreps $A_{g}$ and $A_{u}$, respectively.
However, to keep the comparison with \emph{trans}-polyacetylene transparent,
we will refer to them as $A_{g}$ and $B_{u}$ irreps.

The correlated calculations on the oligomers PDPA-$n$ were performed
using the P-P-P model Hamiltonian \begin{equation}
H=H_{C}+H_{P}+H_{CP}+H_{ee},\label{eq-ham}\end{equation}
 where $H_{C}$ and $H_{P}$ are the one-electron Hamiltonians for
the carbon atoms located on $trans$-polyacetylene backbone (chain),
and the phenyl groups, respectively, $H_{CP}$ is the one-electron
hopping between the chain and the phenyl units and H$_{ee}$ depicts
the electron-electron repulsion. The individual terms can now be written
as,\begin{subequations}
\label{allequations}

\begin{equation}
H_C = -\sum_{\langle k,k' \rangle,M} (t_0 - (-1)^M \Delta t) 
B_{k,k';M,M+1},
  \label{eq-h1}
\end{equation}
\begin{equation}
H_P=-t_0 \sum_{\langle\mu,\nu\rangle,M} B_{\mu,\nu;M,M}, \label{eq-h2}
\end{equation}
and 

\begin{equation}
H_{CP}= -t_{\perp} \sum_{\langle k,\mu \rangle,M} B_{k,\mu;M,M,}. \label{eq-h3}
\end{equation}
\end{subequations}

\begin{eqnarray}
H_{ee} & = & U\sum_{i}n_{i\uparrow}n_{i\downarrow}\nonumber \\
 &  & +\frac{1}{2}\sum_{i\neq j}V_{i,j}(n_{i}-1)(n_{j}-1)\label{eq-hee}\end{eqnarray}

In the equation above, $k$, $k'$ are carbon atoms on the polyene
backbone, $\mu,\nu$ are carbon atoms located on the phenyl groups,
while $i$ and $j$ represent all the atoms of the oligomer. $M$
is a unit consisting of a phenyl group and a polyene carbon, $\langle...\rangle$
implies nearest neighbors, and $B_{i,j;M,M'}=\sum_{\sigma}(c_{i,M,\sigma}^{\dagger}c_{j,M',\sigma}+h.c.)$.
Matrix elements $t_{0}$, and $t_{\perp}$ depict one-electron hops.
In $H_{C}$, $\Delta t$ is the bond alternation parameter arising
due to electron-phonon coupling. In $H_{CP}$, the sum over $\mu$
is restricted to atoms of the phenyl groups that are directly bonded
to backbone carbon atoms. There is a strong possibility that due to
the closeness of the phenyl rings in the adjacent unit cells, there
will be nonzero hopping between them, giving rise to a term $H_{PP}$
in the Hamiltonian above. However, in our earlier study,\cite{shukla2}
we explored the influence of this coupling on the linear optics of
these materials, and found it to have negligible influence. Therefore,
in the present study, we are ignoring the phenyl-phenyl coupling.

As far as the values of the hopping matrix elements are concerned,
we took $t_{0}=2.4$ eV, while it is imperative to take a smaller
value for $t_{\perp}$, because of the twist in the corresponding
bond owing to the steric hindrance mentioned above. We concluded that
for a phenyl group rotation of 30 degrees, the maximum possible value
of $t_{\perp}$ can be 1.4 eV.~\cite{shukla1} Bond alternation parameter
$\Delta t$= 0.168 eV chosen for the backbone was consistent with
the value usually chosen in the P-P-P model calculations performed
on polyenes.

The Coulomb interactions are parameterized according to the Ohno relationship
\cite{ohno}, \begin{equation}
V_{i,j}=U/\kappa_{i,j}(1+0.6117R_{i,j}^{2})^{1/2}\;\mbox{,}\label{eq-ohno}\end{equation}

where, $\kappa_{i,j}$ depicts the dielectric constant of the system
which can simulate the effects of screening, $U$ is the on-site repulsion
term, and $R_{i,j}$ is the distance in \AA ~ between the $i$th
carbon and the the $j$th carbon. The Ohno parameterization initially
was carried out for small molecules, and, therefore, it is possible
that the Coulomb parameters for the polymeric samples could be somewhat
smaller due to interchain screening effects.\cite{chandross} In various
calculations performed on phenylene-based conjugated polymers including
PDPA's\cite{shukla1,shukla2,shukla3,shukla-tpa,shukla-thg,shukla-ppv,hng-ppv},
we have noticed that {}``screened parameters'' with $U=8.0$ eV
and $\kappa_{i,i}=1.0$, and $\kappa_{i,j}=2.0$, otherwise, proposed
by Chandross and Mazumdar~\cite{chandross}, lead to much better
agreement with the experiments, as compared to the ``standard parameters''
with $U=11.13$ and $\kappa_{i,j}=1.0$, proposed originally by Ohno.
Therefore, in the present work we compare the results of our ``screened
parameter'' based calculations with the experiments, while the results
of our ``standard parameter'' based calculations are compared with
similar calculations performed on corresponding polyenes. This is
mainly because screened parameters are not suitable for polyenes.
Moreover, comparison between oligo-PDPA's and polyenes, based upon
calculations performed with the same set of PPP paramters, will help
us appreciate the influence of phenyl substitution on the properties
of PDPA's. 

In all the calculations, C-C bond length of 1.4 \AA ~ was used for
the phenyl rings. In polyenes and PDPA's, along the backbone the single
bonds and the double bonds were taken to be 1.45 \AA~ and 1.35 \AA,
respectively. The bond connecting the backbone to the substituent
phenyl rings was taken to be 1.40 \AA. 

The starting point of the correlated calculations for various oligomers
were the restricted Hartree-Fock (HF) calculations, using the P-P-P
Hamiltonian. The many-body effects beyond HF were computed using different
levels of the configuration interaction (CI) method, namely, quadruples-CI
(QCI), and the multi-reference singles-doubles CI (MRSDCI). Details
of these CI-based many-body procedures have been presented in our
earlier works.\cite{shukla2,shukla-ppv,shukla-tpa,shukla-thg} 

Since the number of $\pi$ electrons in oligo-PDPA's is quite large
because of the large unit cell, it is not possible to include all
the orbitals in the correlated calculations. Therefore, it is imperative
to reduce the number of degrees of freedom by removing some orbitals
from the many-body calculations. In order to achieve that, for each
oligomer we first decided as to which occupied and the virtual orbitals
will be active in the many-body calculations based upon: (a) their
single-particle HF energies with respect to the location of the Fermi
level, and (b) charge distribution of various orbitals with respect
to the chain/phenylene-based atoms, which was quantified by contribution
of the chain-based carbon atoms to the normalization of a given orbital.
Because of the particle-hole symmetry in the problem, the numbers
of active occupied and virtual orbitals were taken to be identical
to each other, with the occupied and virtual orbitals being particle-hole
symmetric. The inactive occupied orbitals were held frozen during
the CI calculations, while the inactive virtual orbitals were simply
deleted from the list of orbitals. When we present the CI results
on various oligo-PDPA's, we will also identify the list of active
orbitals. From the CI calculations, we obtain the eigenfunctions and
eigenvalues corresponding to the correlated ground and excited states
of various oligomers. Using the many-body wave functions, we compute
the matrix elements of the dipole operator amongst different states
and use them, along with the energies of the excited states, to compute
various PA spectra.

\section{Results and Discussion}

\label{results}

In this section we will first briefly discuss the main features of
the experimental PA spectrum obtained by Korovyanko \emph{et al}.,\cite{koro}
and then present and discuss the theoretical results obtained in our
calculations.

\subsection{Experimental Results}

First we briefly review the quantitative aspects of the experimental
results of Korovyanko \emph{et al.\cite{koro}} who measured the PA
spectra of PDPA from $1B_{u}$ and $2A_{g}$ excited states. In $1B_{u}$
PA spectrum they identified two PA bands, the first one of which was
at 1.1 eV called PA1, and the second one at 2.0 eV referred to as
PA2. In the $2A_{g}$ spectrum they identified only one PA band called
PA$_{g}$ located at 1.7 eV. By dipole selection rules it is clear
that the states leading to peaks in $1B_{u}$ spectrum will be of
$A_{g}$ type, while those in the $2A_{g}$ spectrum will be of $B_{u}$
type. 

Another aspect of the PA spectra that we will keep in mind while comparing
the theory to the experimental data, is its connection to nonlinear
optical properties. For example, it is well known that nonlinear optical
properties of conjugated polymers are determined by a small number
of excited states namely $1B_{u}$, $mA_{g}$, and $nB_{u}$.\cite{dixit,sumit}
Here $mA_{g}$ is an excited state which has large dipole coupling
to the $1B_{u}$ state and is visible in TPA and THG spectra, while
$nB_{u}$ is a state with large dipole coupling to the $mA_{g}$ state,
and is visible in the THG spectra of conjugated polymers.\cite{dixit,sumit}
Thus it is obvious that the $mA_{g}$ state should contribute significantly
to the $1B_{u}$ PA spectra. Whether $nB_{u}$ will contribute to
the $2A_{g}$ PA spectrum will depend on, whether or not, it has a
large dipole coupling with the $2A_{g}$ state. In a recent theoretical
analysis of the $1B_{u}$ PA spectra of PPV which also has two prominent
bands labeled PA1 and PA2, we indeed found that the first and the
most intense peak (PA1) contributing to the spectra was indeed the
$mA_{g}$ state of the polymer.\cite{hng-ppv}

\subsection{Theoretical Results}

We have performed correlated calculations of the PA spectrum of oligomers
PDPA-5 and PDPA-10. The reasons behind choosing these oligomers is
that the conjugation lengths of the oligomers under experimental conditions
is believed to be short.\cite{gontia2} Resonant Raman scattering
based studies in films of PDPA-$n$Bu have indicated that the mean
conjugation length is seven repeat units,\cite{fujii2} while in the
solution phase it is believed to be close to five repeat units.\cite{gontia2}
Therefore, we believe that our choice of the oligomers PDPA-5 and
PDPA-10 is well-suited for comparisons with experiments.

\begin{figure}

\caption{The occupied energy levels of PDPA-5 obtained from the Hartree-Fock
calculations performed using the screened parameters. As indicated,
the Fermi energy has been shifted to zero. On the left of each level
there are two numbers, first of which is the energy of the level in
eV, and the second one is the charge distribution of the orbital quantified
as the contribution of chain carbon atoms to the orbital normalization.
Contribution of the carbon atoms on the phenyl rings, to the orbital
charge distribution can be obtained by subtracting this number from
one. On the right of each level is the location of the level with
respect to the HOMO (H). The thick line indicated as $l$ represents
the band composed of localized orbitals of the phenyl rings. Because
of the particle-hole symmetry of the half-filled PPP model, virtual
levels will be mirror symmetric with respect to the Fermi level. Additionally,
virtual orbital $L+i$ will have the same charge distribution as the
occupied orbital $H-i$. }

\includegraphics[%
  scale=0.7]{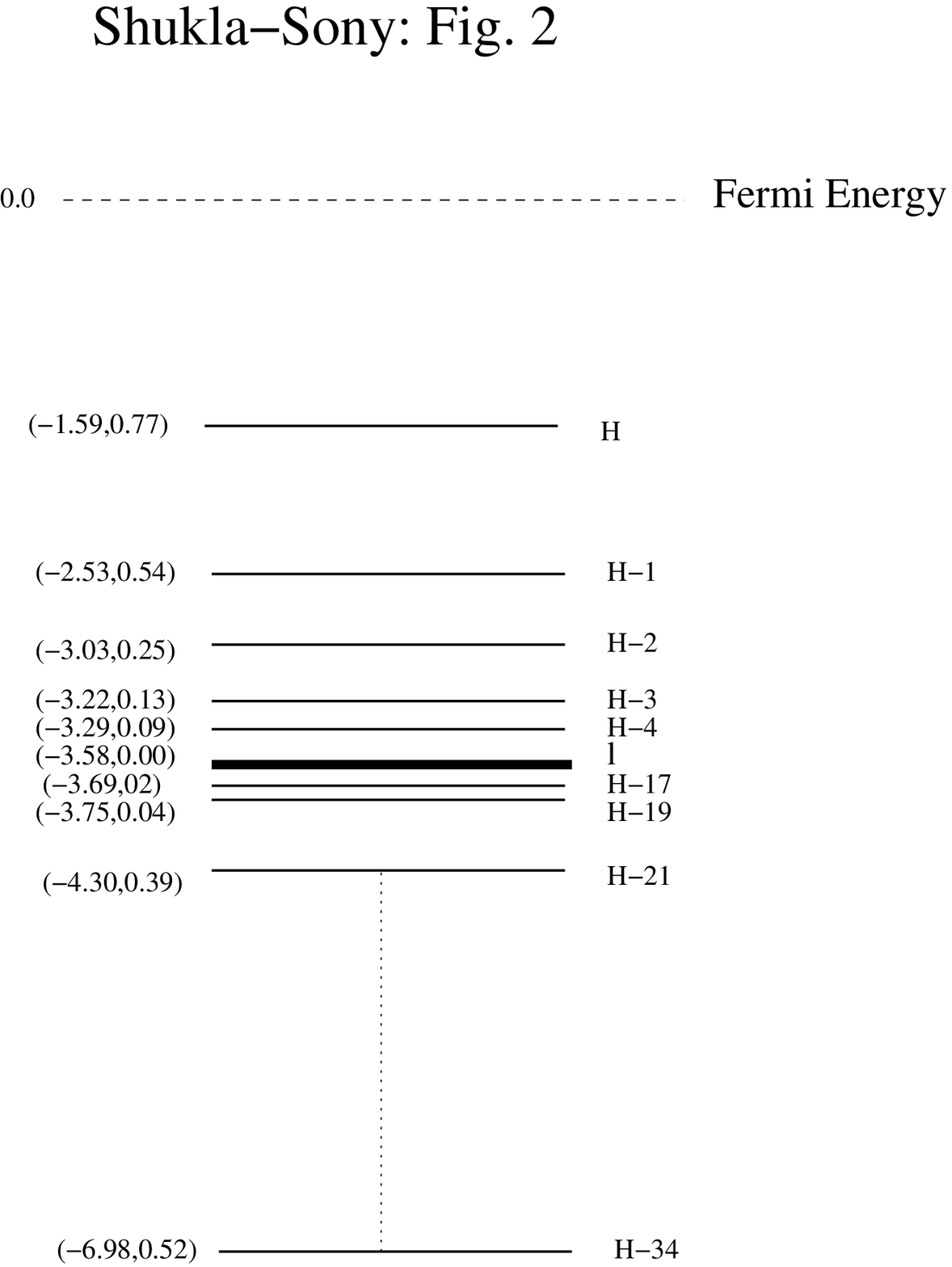}\label{fig-level}
\end{figure}

Before presenting our results, we would like to briefly describe the
type of correlated calculations performed, and the approximations
involved. The correlated calculations performed were of two types:
(a) QCI and (b) MRSDCI. Given the large number of $\pi$ electrons
in these systems (fourteen $\pi$ electrons/cell), QCI or the MRSDCI
methods are not feasible for them if all the orbitals of the system
are retained in the calculations. Since it is intuitively clear that
the low-lying excited states of the system will be determined by orbitals
closest to the Fermi level which have large charge distribution of
backbone carbon atoms, in the QCI calculations we decided to include
only these orbitals. Therefore, for PDPA-$n$ we included $n$ occupied,
and $n$ virtual orbitals closest to the Fermi level in the QCI calculations.
Remaining occupied orbitals were frozen and virtual orbitals were
deleted as explained in section \ref{method}. The charge distribution
of these $n$ orbitals is as follows: HOMO/LUMO ($H/L$ for short)
orbitals have the largest charge distribution on the chain-based atoms
with this quantity decreasing monotoncially for $H-i/L+i$ orbitals
with increasing $i$ ($\leq n)$. Levels located next are $n$ $l/l^{*}$
orbitals which have zero contribution from the chain atoms, followed
by other orbitals. The orbital energy diagram for PDPA-5 is presented
in Fig. \ref{fig-level}. The computational effort associated with
the QCI calculations on PDPA-$n$ is, therefore, same as that needed
for a polyene with $n$ double bonds. Although for PDPA-$10$, it
leads to Hilbert space dimensions in excess of one million, however,
using the approach reported in our earlier works\cite{shukla-ppv,shukla2},
we managed to obtain low-lying excited states for this set of calculations.
We do believe that, despite the orbital truncation, the QCI calculations
should describe the lowest of the excited states reasonably well. 

In order to improve upon the aforesaid orbital truncation scheme,
we augmented the orbital set by including orbitals with large charge
distribution on the phenylene-based carbon atoms. Phenylene-based
orbitals are of two-types: (a) delocalized orbitals ($d/d^{*}$) with
significant population on all six phenyl-based carbon atoms and (b)
localized orbitals ($l/l^{*}$) with nonzero population only on four
phenyl-based carbon atoms. It is intuitively clear that $l/l^{*}$
type orbitals will participate more in the excitations localized on
the phenyl rings, because these orbitals will have negligible population
on the phenyl-based carbon atom connected to the back bone (cf. Fig.
\ref{fig-pdpa}). On the other hand $d/d^{*}$ orbitals in PDPA's
will be involved in excitations extending from backbone to the phenyl
rings. Since it is these type of excitations which are next in energy,
we augmented the orbital set by also including the phenyl-based $d/d^{*}$
orbitals. However, with this augmented orbital set, only MRSDCI calculations
are feasible, and those too on oligomers with comparatively smaller
conjugation length. But, given the small conjugation lengths in the
experimental samples discussed above, we believe that this is not
a severe limitation. Thus, these MRSDCI calculations were performed
on PDPA-5 using thirty orbitals (15 occupied + 15 virtual) in all,
ten of which were chain-based orbitals closest to the Fermi level
(also included in the QCI calculations mentioned above), while remaining
twenty were from the $d/d^{*}$ phenyl-based orbital set mentioned
above. In the MRSDCI calculations, we used 25 reference configurations
for the $A_{g}$-type states, and 24 for the $B_{u}$-type states
leading to CI matrices of dimensions close to half-a-million both
for the $A_{g}$ and $B_{u}$ manifolds. 

Thus, to summarize, we present QCI calculations performed on PDPA-5
and PDPA-10 utilizing restricted orbital sets, and MRSDCI calculations
on PDPA-5 utilizing an augmented orbital set. QCI calculations were
performed with orbitals closest to the Fermi level, several of which
had large charge distribution on carbon atoms based on the backbone.
MRSDCI calculations were performed with an orbital set which, in addition
to the orbitals used in the QCI calculations, also contained $d/d^{*}$
with large charges on the carbon atoms based on phenyl rings. Thus,
the results of QCI and MRSDCI calculations should be in good agreement
with each other for those excited states, which can be described mainly
in terms of orbitals based on backbone atoms. While the excited states
in which electrons get significantly delocalized onto the phenyl rings,
will be described better by MRSDCI calculations. Thus, by comparing
the results of these two types of calculations for PDPA-5 with each
other, we can assess as to where the results of the two approaches
begin to diverge from each other. This will also let us form an opinion
about the results obtained for PDPA-10, for which no MRSDCI results
are available. It is with this information about the uncertainties
associated with various calculations, we present our theoretical results
and compare them to the experimental ones, in the next section.

\subsubsection{$1B_{u}$ PA Spectrum}

\begin{figure}

\caption{PA spectrum of PDPA-5 from its $1B_{u}$ state, computed using: (a)
MRSDCI method, and (b) QCI method. First three peaks have been labeled
and discussed in the text. A linewidth of 0.15 eV was assumed.}

\includegraphics[%
  scale=0.7,
  angle=-90]{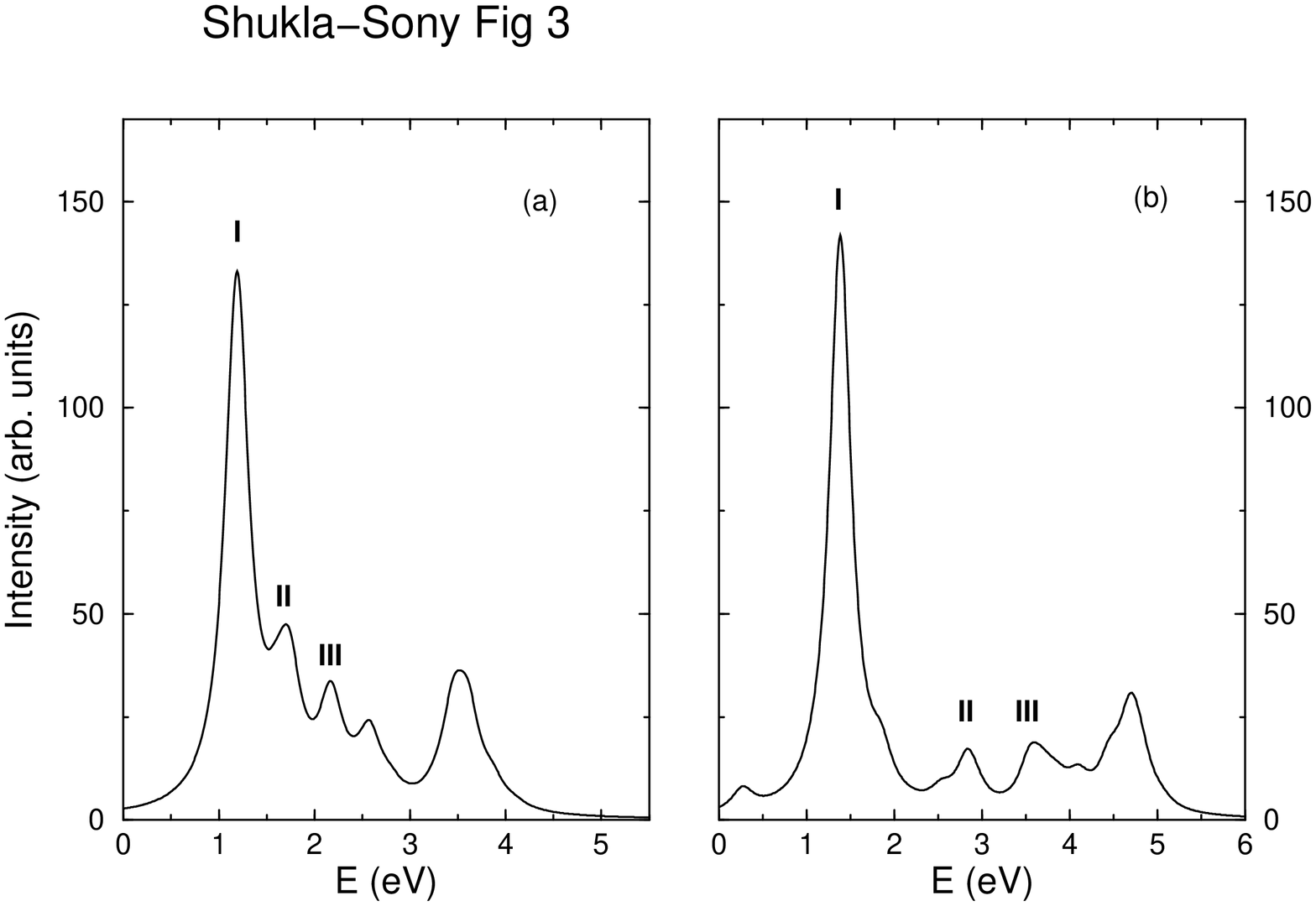}\label{pdpa5-1bu}
\end{figure}

\begin{figure}

\caption{PA spectrum of PDPA-10 from its $1B_{u}$ state, computed using the
QCI method. Only the peaks in the experimental energy region have
been labeled. A linewidth of 0.15 eV was assumed.}

\includegraphics[%
  width=8cm,
  angle=-90]{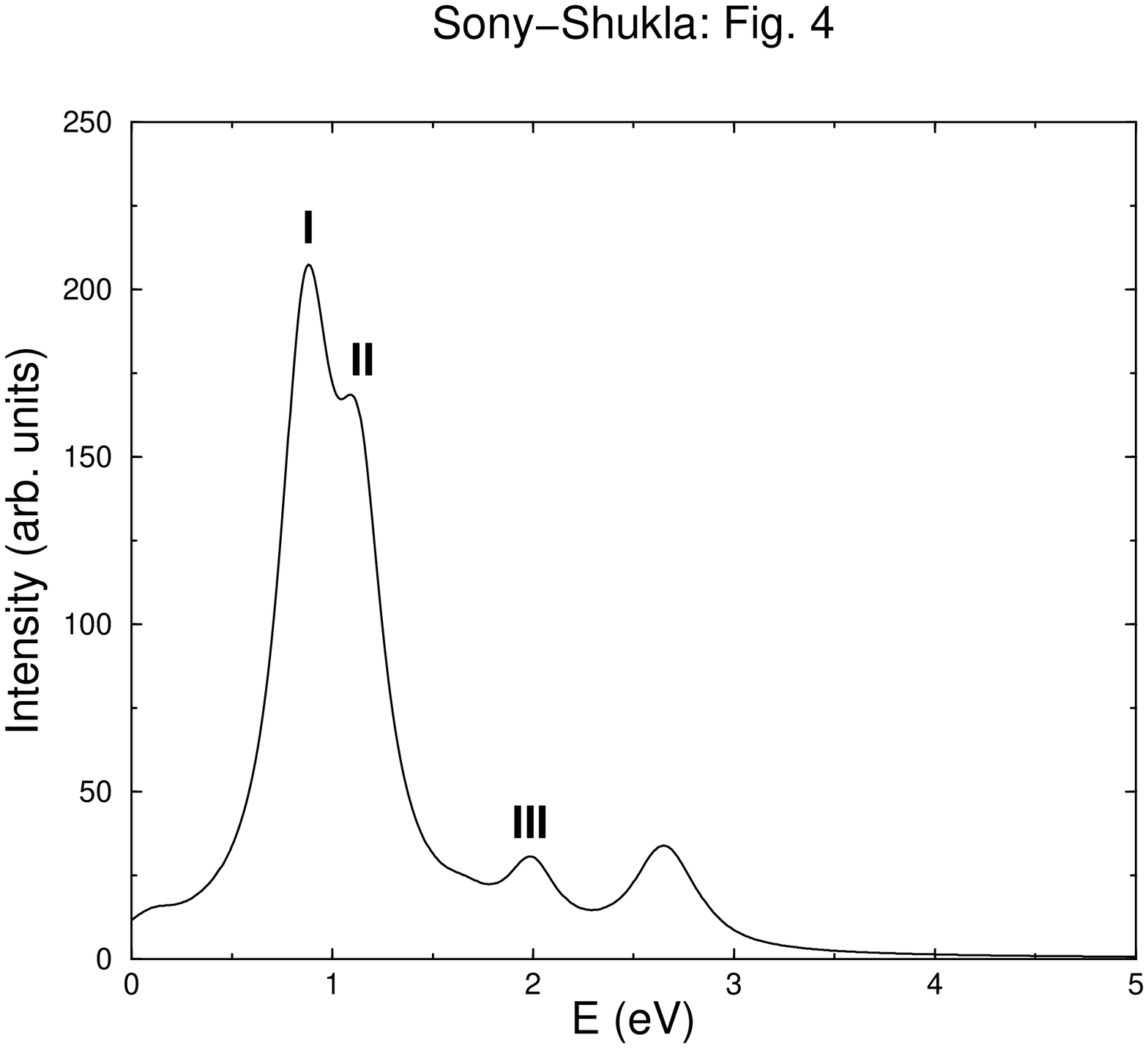}\label{pdpa10-1bu-qci}
\end{figure}
The spectrum from $1B_{u}$ excited state of PDPA-5 calculated using
QCI and MRSDCI methods, and the screened parameters, is shown in Fig.
\ref{pdpa5-1bu}. For PDPA-10 the same spectrum computed using the
QCI method is given in Fig. \ref{pdpa10-1bu-qci}. In Tables \ref{tab-pdpa5-1bu}
and \ref{tab-pdpa10-qci-1bu} we present energies and wave functions
of the excited states, contributing to the spectra of Figures \ref{pdpa5-1bu}
and \ref{pdpa10-1bu-qci}. Before discussing the spectra in detail,
we identify their broad features. It is clear from the figures that
each spectrum starts with a very intense peak followed by a number
of weaker features. Upon comparing the calculated intensity profile
with the experimental one,\cite{koro} we find general agreement between
the two. However, in the experimental spectrum the intensity of feature
PA2 is not as drastically smaller compared to PA1 as in case of theoretical
spectrum. 

Upon examining the theoretical results with experimental feature PA1
in mind, in both the MRSDCI and QCI spectra corresponding to PDPA-5,
only one excited state contributes to the first feature. In PDPA-10
spectrum, the first absorption band has one intense peak (I) followed
by a closely spaced shoulder (II), contributions to which come from
three rather closely spaced excited states. Thus for PDPA-10 (Fig.
\ref{pdpa10-1bu-qci}), we interpret both I and II as part of the
same peak, to be compared to the experimental band PA1. Upon examining
the many-particle wave functions of the excited states contributing
to the first peak in various calculations (tables \ref{tab-pdpa5-1bu}
and \ref{tab-pdpa10-qci-1bu}) we conclude that the configurations
contributing to them are essentially identical in all the cases. Since
this conclusion is based upon calculations of various types (QCI,
MRSDCI), and on different oligomers (PDPA-5, PDPA-10), we conclude
that these results are universally valid for PDPA's. Comparing these
states to ones contributing resonant peaks of the TPA and THG spectra
of oligo-PDPA's calculated by us earlier\cite{shukla-tpa,shukla-thg},
we conclude that these states correspond to nothing but the $mA_{g}$
states of the corresponding oligo-PDPA. Recalling that $mA_{g}$ is
the state with a very strong dipole coupling to the $1B_{u}$ state
which makes its presence prominent in both the TPA and THG spectra
of several conjugated polymers,\cite{sumit,hng-ppv} it should not
at all be surprising that it leads to an intense peak in their $1B_{u}$
PA spectrum as well. Indeed, in a recent calculation performed on
oligomers of PPP and PPV,\cite{hng-ppv} we identified the first feature
of their $1B_{u}$ PA spectra with their $mA_{g}$ states. Upon examining
the dipole coupling of $mA_{g}$ states of oligo-PDPA's to their $1B_{u}$
states, we find: (a) this to be mainly directed along the longitudinal
direction, and (b) that in terms of magnitude the coupling of this
state to the $1B_{u}$ state to be the largest among all other $A_{g}$
type excited states. 

Regarding the quantitative aspects of our calculations, it is obvious
from table \ref{tab-pdpa5-1bu} that for PDPA-5, the peak position
of PA1 obtained from the QCI calculations (1.38 eV) is in good agreement
(within 0.2 eV) of the one computed from the MRSDCI calculation (1.19
eV). This result is consistent with the fact that for PDPA-5, wave
functions of the states contributing to PA1 obtained from QCI and
MRSDCI calculations are in excellent agreement with each other, and
consist mainly of excitations close to the Fermi level (cf. table
\ref{tab-pdpa5-1bu}). The calculated peak positions of PA1 for various
oligomers are presented in table \ref{tab-sum}, along with the corresponding
experimental value. We note that results of MRSDCI calculation for
PDPA-5 (1.19 eV) appear to be in the best agreement with the experiments
(1.1---1.2 eV).

For features beyond PA1 for PDPA-5, the results of QCI and MRSDCI
calculations begin to diverge from each other both for the wave functions
of the excited states, as also their peak positions. As far as experimental
feature PA2 (at 2.0 eV) is concerned, we find that none of the peaks
in the QCI spectrum of PDPA-5 are close to that. However, in the MRSDCI
spectrum of PDPA-5, feature II located at 1.72 eV, and feature III
located at 2.17 eV, could be considered as possible candidates. Upon
examining the MRSDCI wave functions of PDPA-5 in table \ref{tab-pdpa5-1bu}
of states $6A_{g}$ (peak II) and $9A_{g}$ (peak III), we find that
they predominantly consist of singly excited configurations. In case
of peak II, these singly excited configurations have almost equal
contributions from low-lying excitations as well as high-lying excitations,
involving chain and phenylene based delocalized orbitals. However,
state $9A_{g}$ (peak III) mainly consists of singly excited high-lying
configurations among the chain based orbitals and the phenylene based
orbitals. Looking at the polarization of the photons involved, the
computed dipole connecting the $1B_{u}$ state to states $6A_{g}$
suggests mixed polarization, with $y-$ component (transverse) being
more intense. The corresponding dipole moment connecting $1B_{u}$
state to the $9A_{g}$ state, on the other hand, suggests mainly an
$x-$ polarized (longitudinal) transition. 

In the QCI spectrum of PDPA-10, feature III which is located exactly
at 2.0 eV is can also be considered a very good candidate for PA2.
The many particle wave function of the state involved ($11A_{g}$in
table \ref{tab-pdpa10-qci-1bu}) is a mixture of higher energy singly
and doubly excited configurations. As far as transition dipoles are
concerned, we find it to be an almost equal mixture of $x-$ and $y-$
components. Thus, as far as the polarization of the photon, and the
nature of excited states involved in the feature PA2 we have a disagreement
between our MRSDCI results on PDPA-5, and QCI results on PDPA-10.
In order to resolve this issue, it will be of interest to perform
experiments on oriented samples, and measure the polarization of the
photons contributing to the feature PA2.  

\begin{table}

\caption{Excited states contributing to the $1B_{u}$ PA spectrum of PDPA-5
computed using the QCI and the MRSDCI methods corresponding to the
spectra of Fig. \ref{pdpa5-1bu}. The heading Wave Function lists
the most important configurations contributing to the many-body wave
function, and their coefficients, as per our adopted notation.\cite{notation}}

\begin{tabular}{|c|c|c|c|c|}
\hline 
Method&
Peak&
State&
 Peak Position (eV) &
Wave Function\tabularnewline
\hline
\hline 
QCI&
I&
$4A_{g}(mA_{g})$&
1.38&
$|H\rightarrow L+1\rangle+c.c.(0.37)$\tabularnewline
&
&
&
&
$|H\rightarrow L;H\rightarrow L\rangle(0.74)$\tabularnewline
\hline 
MRSDCI&
I&
$4A_{g}(mA_{g})$&
1.19&
$|H\rightarrow L+1\rangle+c.c.(0.41)$\tabularnewline
&
&
&
&
$|H\rightarrow L;H\rightarrow L\rangle(0.58)$\tabularnewline
\hline 
QCI&
II&
9$A_{g}$&
2.83&
$|H\rightarrow L+1;H-1\rightarrow L\rangle+(0.47)$\tabularnewline
&
&
&
&
$|H\rightarrow L;H\rightarrow L\rangle+c.c.(0.36)$\tabularnewline
&
&
&
&
$|H\rightarrow L;H-4\rightarrow L\rangle+c.c.(0.22)$\tabularnewline
&
&
&
&
$H\rightarrow L;H-2\rightarrow L\rangle+c.c.(0.21)$\tabularnewline
\hline 
MRSDCI&
II&
$6A_{g}$&
1.72&
$|H\rightarrow L+19\rangle+c.c.\:(0.38)$\tabularnewline
&
&
&
&
$|H-2\rightarrow L+1\rangle+c.c.(0.30)$\tabularnewline
&
&
&
&
$|H\rightarrow L+17\rangle+c.c.(0.26)$\tabularnewline
\hline 
QCI&
III&
12$A_{g}$&
3.56&
$|H\rightarrow L+1;H-1\rightarrow L\rangle+c.c.(0.51)$\tabularnewline
&
&
&
&
$|H\rightarrow L;H-2\rightarrow L\rangle+c.c(0.38)$\tabularnewline
&
&
&
&
$|H\rightarrow L;H-3\rightarrow L\rangle+c.c(0.32)$\tabularnewline
\hline
MRSDCI&
III&
$9A_{g}$&
2.17&
$|H\rightarrow L+21\rangle+c.c.(0.57)$\tabularnewline
&
&
&
&
$|H\rightarrow L;H\rightarrow L\rangle(0.18)$\tabularnewline
\hline
\end{tabular}

\label{tab-pdpa5-1bu}
\end{table}

\begin{table}

\caption{Excited states contributing to the $1B_{u}$ PA spectrum of PDPA-10,
computed using the QCI method, and presented in Fig. \ref{pdpa10-1bu-qci}.
Rest of the information is same as given in the caption of table \ref{tab-pdpa5-1bu}. }

\begin{tabular}{|c|c|c|c|}
\hline 
Peak&
State&
Peak Position (eV)&
Wave Function\tabularnewline
\hline
I&
$3A_{g}(mA_{g})$&
0.81&
$|H\rightarrow L;H\rightarrow L\rangle(0.37)$\tabularnewline
&
&
&
$|H\rightarrow L+3\rangle+c.c.(0.31)$\tabularnewline
&
&
&
$|H\rightarrow L+1\rangle+c.c.(0.29)$\tabularnewline
&
&
&
$|H\rightarrow L+1;H\rightarrow L+1\rangle+c.c.(0.29)$\tabularnewline
&
$4A_{g}(mA_{g})$&
0.87&
|$H\rightarrow L+3\rangle+c.c.(0.42)$\tabularnewline
&
&
&
$|H\rightarrow L+1\rangle+c.c.(0.27)$\tabularnewline
&
&
&
$|H\rightarrow L;H\rightarrow L\rangle(0.33)$\tabularnewline
&
&
&
$|H-1\rightarrow L;H\rightarrow L+1\rangle(0.31)$\tabularnewline
&
&
&
$|H\rightarrow L;H\rightarrow L+2\rangle+c.c.(0.21)$\tabularnewline
\hline 
II&
$5A_{g}$($mA_{g})$&
1.12&
$|H\rightarrow L;H\rightarrow L\rangle(0.38)$\tabularnewline
&
&
&
$|H-1\rightarrow L;H\rightarrow L+1\rangle(0.33)$\tabularnewline
&
&
&
$|H\rightarrow L;H\rightarrow L+2\rangle+c.c.(0.29)$\tabularnewline
&
&
&
$|H-1\rightarrow L+2\rangle+c.c.(0.26)$\tabularnewline
&
&
&
$|H\rightarrow L+1\rangle+c.c.(0.20)$\tabularnewline
\hline 
III&
$11A_{g}$&
1.99&
$|H\rightarrow L;H-2\rightarrow L+2\rangle(0.30)$\tabularnewline
&
&
&
$|H\rightarrow L+3;H-1\rightarrow L\rangle+c.c.(0.21)$\tabularnewline
&
&
&
$|H-2\rightarrow L+3\rangle+c.c.(0.26)$\tabularnewline
\hline
\end{tabular}\label{tab-pdpa10-qci-1bu}
\end{table}

\begin{figure}

\caption{Comparison of $1B_{u}$ PA spectra of oligo-PDPA's with polyenes
of the same conjugation length: (a) spectra of PDPA-5 (solid line)
and five double-bond polyene (dashed line), and (b) spectra of PDPA-10
(solid line) and ten double-bond polyene (dashed line). All the spectra
were computed using the QCI method and the standard parameters in
the PPP Hamiltonian. For PDPA-$n$, QCI calculations were performed
with $n$ occupied/virtual orbitals closest to Fermi level. A common
linewidth of 0.15 eV was used to compute the spectra.}

\includegraphics[%
  scale=0.7,
  angle=-90]{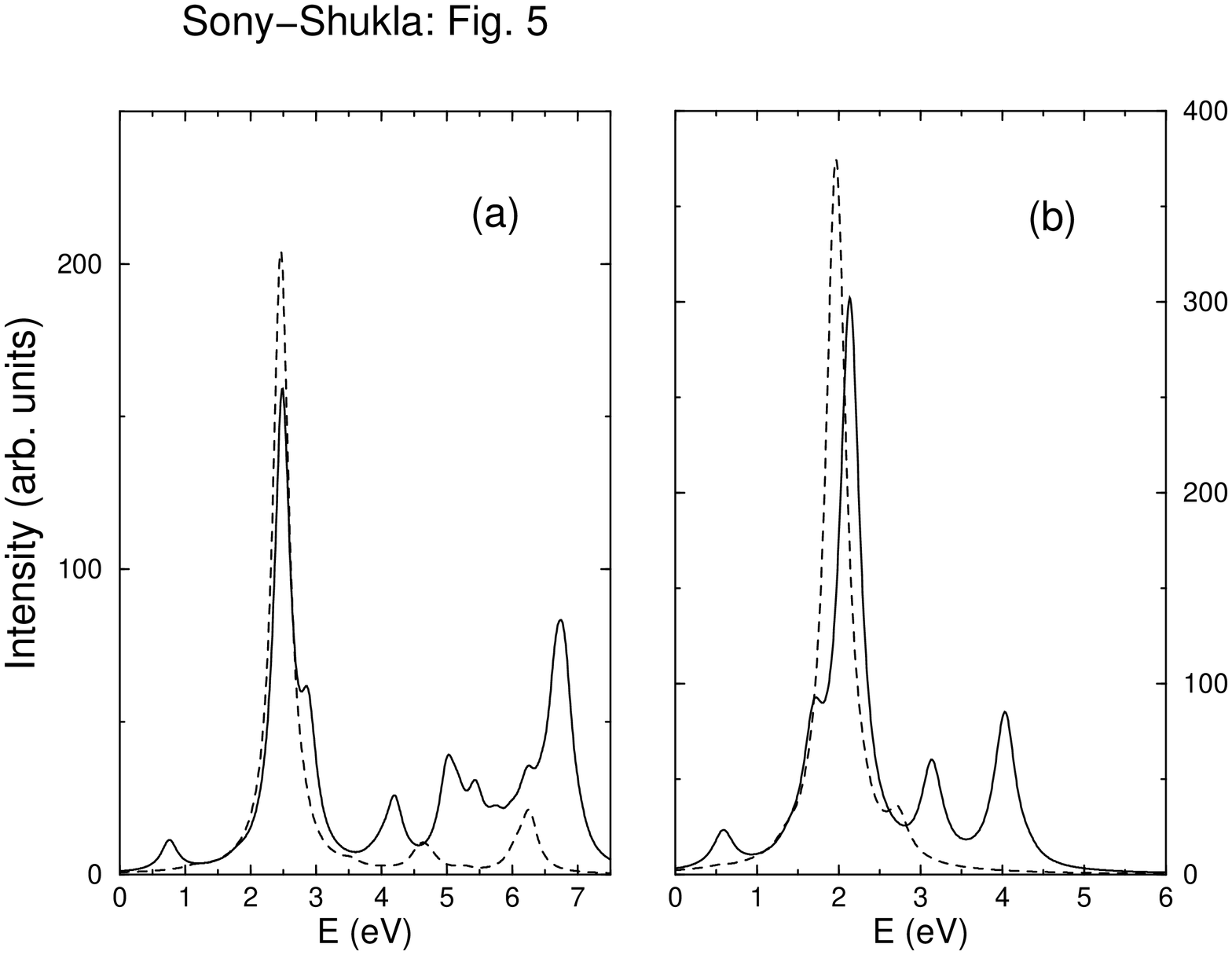}\label{tpa-pa-1bu}
\end{figure}

Next we compare the $1B_{u}$ PA spectra of oligo-PDPA's with those
of corresponding polyenes. Both for polyenes and the PDPA's, we use
the standard parameters in the PPP Hamiltonian, coupled with the QCI
method. Thus, calculations on oligo-PDPA's were performed with a restricted
set of orbitals as explained earlier, while for polyenes all the orbitals
were used. The results of our calculations are presented in Fig. \ref{tpa-pa-1bu}.
The peak positions for both PDPA-5 and PDPA-10 are blue shifted as
compared to those computed with the screened parameters. For example,
the position of the first peak for PDPA-5 with standard parameters
is 2.5 eV, as compared to 1.38 eV obtained with the screened parameters
and the 1.2 eV experimental value. Thus, quantitatively speaking,
the PA spectra of oligo-PDPA's computed using the standard parameters
are not in good agreement with the experimental results. From Fig.
\ref{tpa-pa-1bu} we also conclude that the first peak of both oligo-PDPA's
and the corresponding polyenes are both due to the $mA_{g}$ states
of the system concerned. This lends further credence to our earlier
conclusion that $mA_{g}$ state of oligo-PDPA's, which corresponds
to the PA1 feature of the experimental spectrum, is mainly due to
chain-based excitations. Also from Fig. \ref{tpa-pa-1bu} it is obvious
that, unlike oligo-PDPA's, in the polyene PA spectra there is no significant
structure beyond the first peak. This is in excellent agreement with
our earlier conclusion that the experimental feature PA2 is mainly
due to the excitations involving phenylene-based orbitals. 

\subsubsection{$2A_{g}$ PA spectrum}

\begin{figure}

\caption{PA spectrum of PDPA-5 from its $2A_{g}$ excited state, computed
using: (a) MRSDCI method, and (b) QCI method. A linewidth of 0.15
eV was assumed. Feature I corresponds to the PA$_{g}$ feature of
the experiment.\cite{koro}Additionally, we have labeled and discussed
(see text) the next peak (feature II) of the spectra, with the possibility
of it being detected in future experiments.}

\includegraphics[%
  scale=0.7,
  angle=-90]{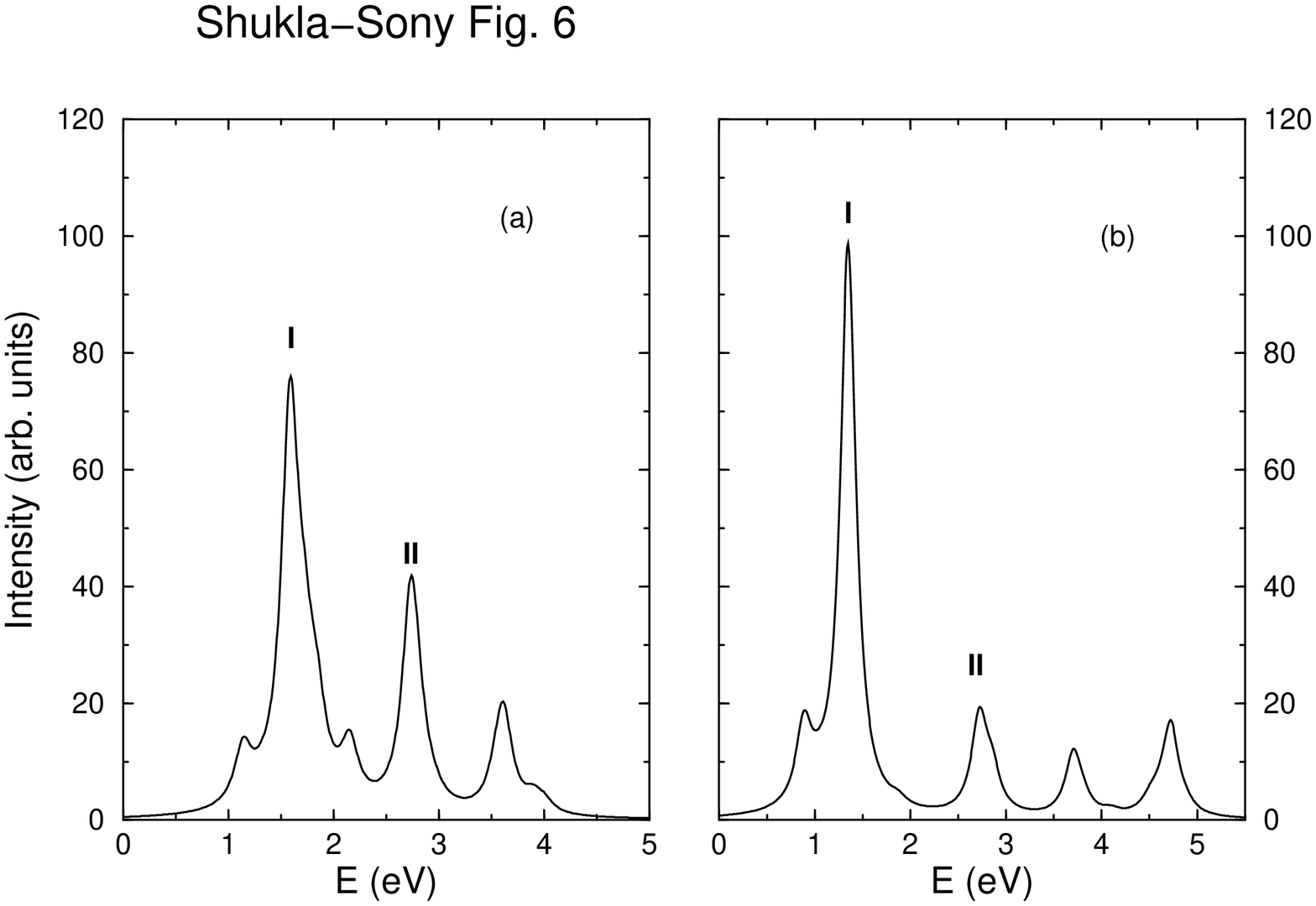}\label{pdpa5--w15-2ag}
\end{figure}

\begin{figure}

\caption{PA spectrum of PDPA-10 from its $2A_{g}$ excited state, computed
using the QCI method. Only the peaks in the experimental region have
been labeled. A linewidth of 0.15 eV was assumed.}

\includegraphics[%
  width=8cm,
  angle=-90]{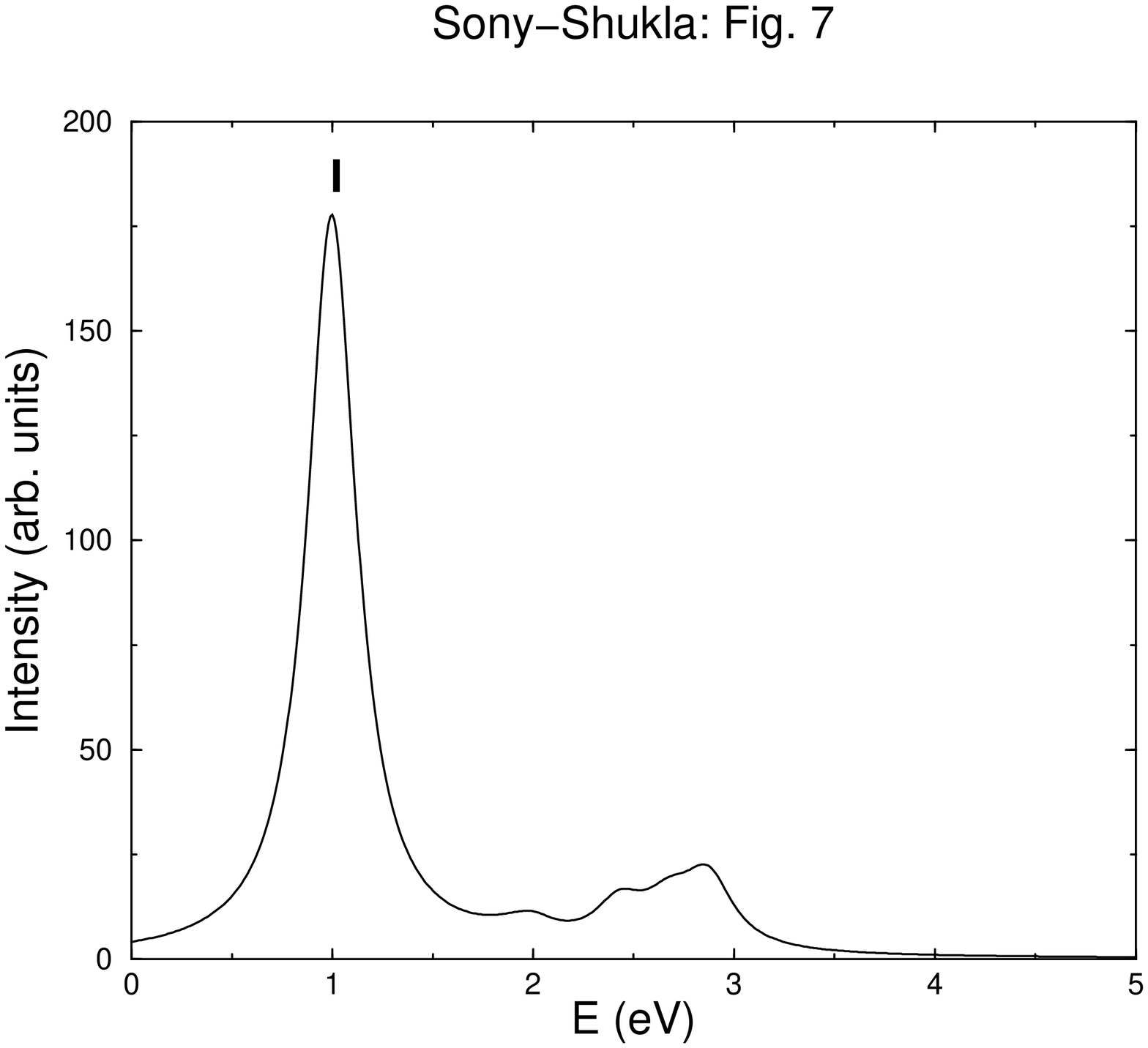}\label{pdpa10-qci-w15-2ag}
\end{figure}

The PA spectra from the $2A_{g}$ excited state of PDPA-5 computed
using the screened parameters, at the MRSDCI and QCI levels, are shown
in Fig. \ref{pdpa5--w15-2ag}, while that for PDPA-10 computed at
the QCI level is presented in Fig. \ref{pdpa10-qci-w15-2ag}. The
peak positions and the wave functions of the many-body states for
these spectra are listed in tables \ref{tab-pdpa5-2ag} and \ref{tab-pdpa10-qci-2ag}.
It is clear from the theoretical PA$_{g}$ spectra that the intensity
is dominated by just one peak (peak I) with smaller contributions
from a series of weak peaks. This aspect of theoretical spectrum is
in excellent agreement with the experiment which reports only one
peak.\cite{koro} We identify peak I in all the computed spectra as
the $nB_{u}$ state of the polymer. The same $nB_{u}$ state had also
contributed significantly to the longitudinal THG spectra of these
oligomers as reported in our earlier work. \cite{shukla-thg} The
$nB_{u}$ state reported here has a significantly strong $y-$ polarized
dipole coupling to the $1A_{g}$ ground state. However, it has a much
stronger $x-$ polarized dipole coupling to the $2A_{g}$ state. Therefore,
in experiments performed on oriented samples, we predict the PA$_{g}$
feature of oligo-PDPA's to be mainly due to $x$-polarized photons
leading to an absorption from the $2A_{g}$ state to the $nB_{u}$
state. In case of conjugated polymers such as \emph{trans}-polyacetylene,
PPV, and PPP etc.,\cite{sumit,hng-ppv} the $nB_{u}$ state has strong
dipole coupling to the $mA_{g}$ state. However, in PDPA's the state
we have identified as $nB_{u}$ has a very strong dipole coupling
with the $2A_{g}$ state, and a comparatively weaker coupling to the
$mA_{g}$ state. The state with large dipole coupling to the $mA_{g}$
state contributes to the computed THG spectrum of PDPA's, and was
identified earlier by us as the $kB_{u}$ state. \cite{shukla-thg} 

As far as the many-particle wave function of the state $nB_{u}$ corresponding
to the $PA_{g}$ feature is concerned, we have excellent agreement
among various calculations performed on different oligomers. From
tables \ref{tab-pdpa5-2ag} and \ref{tab-pdpa10-qci-2ag} it is obvious
that this state mainly consists of singly excited configuration among
orbitals $H-1$ and $L+1$ . Regarding the calculated peak positions
of PA$_{g}$, for PDPA-5 the MRSDCI value (1.59 eV) differs from the
QCI value (1.35 eV) by 0.24 eV. Thus, the disagreements between the
results of QCI and MRSDCI calculations for PDPA-5 both for PA1 (0.20
eV) and PA$_{g}$ (0.24 eV) are almost similar in size. The only difference
between the MRSDCI and the QCI for PDPA-5 spectra is regarding the
intensity, and, quite expectedly, the nature of the states contributing
to peaks beyond PA$_{g}$ in the spectra. For example, in Fig. \ref{pdpa5--w15-2ag},
peak II in the MRSDCI spectrum of PDPA-5 has considerably larger intensity
as compared to the same peak in the QCI spectrum. Peak II corresponds
to the $kB_{u}$ state for the case of PDPA-5 and has been discussed
earlier. \cite{shukla-thg} The character of the $kB_{u}$ state as
computed in the MRSDCI calculations differs somewhat from that computed
in the QCI calculations in that MRSDCI $kB_{u}$ state has significant
contribution from the singly excited configurations involving phenyl
based orbitals. However, both the MRSDCI and QCI calculations predict
the $2A_{g}-kB_{u}$ optical transition to be mainly $y$-polarized.
Thus, if future experiments on oriented samples are able to probe
features beyond PA$_{g}$ in the $2A_{g}$-PA spectrum of oligo-PDPA's,
the results of our calculations can be tested. %
\begin{table}

\caption{Excited states contributing to the $2A_{g}$ PA spectrum of PDPA-5
computed using the QCI and MRSDCI methods, and presented in Fig. \ref{pdpa5--w15-2ag}.
Rest of the information is same as given in table \ref{tab-pdpa5-1bu}.}

\begin{tabular}{|c|c|c|c|c|}
\hline 
Method&
Peak&
State&
Peak Energy (eV)&
Wave Function\tabularnewline
\hline
\hline 
QCI&
I&
$4B_{u}(nB_{u})$&
1.35&
$|H-1\rightarrow L+1\rangle(0.94)$\tabularnewline
&
&
&
&
$|H\rightarrow L;H\rightarrow L+1\rangle+c.c.(0.14)$\tabularnewline
\hline 
MRSDCI&
I&
$4B_{u}(nB_{u})$&
1.59&
$|H-1\rightarrow L+1\rangle(0.79)$\tabularnewline
&
&
&
&
$|H-16\rightarrow L\rangle+c.c.(0.15)$\tabularnewline
&
&
&
&
$|H\rightarrow L;H-1\rightarrow L\rangle+c.c.(0.13)$\tabularnewline
\hline 
QCI&
II&
$9B_{u}(kB_{u})$&
2.73&
$|H\rightarrow L;H\rightarrow L+1\rangle+c.c.(0.57)$\tabularnewline
&
&
&
&
$|H-1\rightarrow L+1;H-1\rightarrow L\rangle+c.c.(0.25)$\tabularnewline
&
&
&
&
$|H-1\rightarrow L+1\rangle(0.19)$\tabularnewline
\hline 
MRSDCI&
II&
$9B_{u}(kB_{u})$&
2.74&
$|H-1\rightarrow L+19\rangle+c.c.(0.48)$\tabularnewline
&
&
&
&
$|H\rightarrow L;H\rightarrow L+1\rangle+c.c.(0.27)$\tabularnewline
&
&
&
&
$|H-4\rightarrow L+4\rangle(0.20)$\tabularnewline
\hline
\end{tabular}\label{tab-pdpa5-2ag}
\end{table}

\begin{table}

\caption{Excited state contributing to the $2A_{g}$ PA spectrum of PDPA-10,
computed using the QCI method, and presented in Fig. \ref{pdpa10-qci-w15-2ag}.
Rest of the information is same as given in table \ref{tab-pdpa5-1bu}.}

\begin{tabular}{|c|c|c|c|}
\hline 
Peak&
State&
Peak Energy (eV)&
Wave Function\tabularnewline
\hline
I&
$3B_{u}(nB_{u})$&
1.00&
$|H-1\rightarrow L+1\rangle(0.72)$\tabularnewline
&
&
&
$|H\rightarrow L+2\rangle+c.c.(0.38)$\tabularnewline
&
&
&
$|H\rightarrow L;H-1\rightarrow L\rangle+c.c.(0.14)$\tabularnewline
\hline
\end{tabular}\label{tab-pdpa10-qci-2ag}
\end{table}

\begin{figure}

\caption{$2A_{g}$-PA spectra of (a) PDPA-5 (solid line) and five double-bond
polyene (dashed line), and (b) PDPA-10 (solid line) and ten double-bond
polyene (dashed line). Rest of the information is same as in the caption
of Fig. \ref{tpa-pa-1bu}.\label{tpa-pa-2ag}}

\includegraphics[%
  scale=0.7,
  angle=-90]{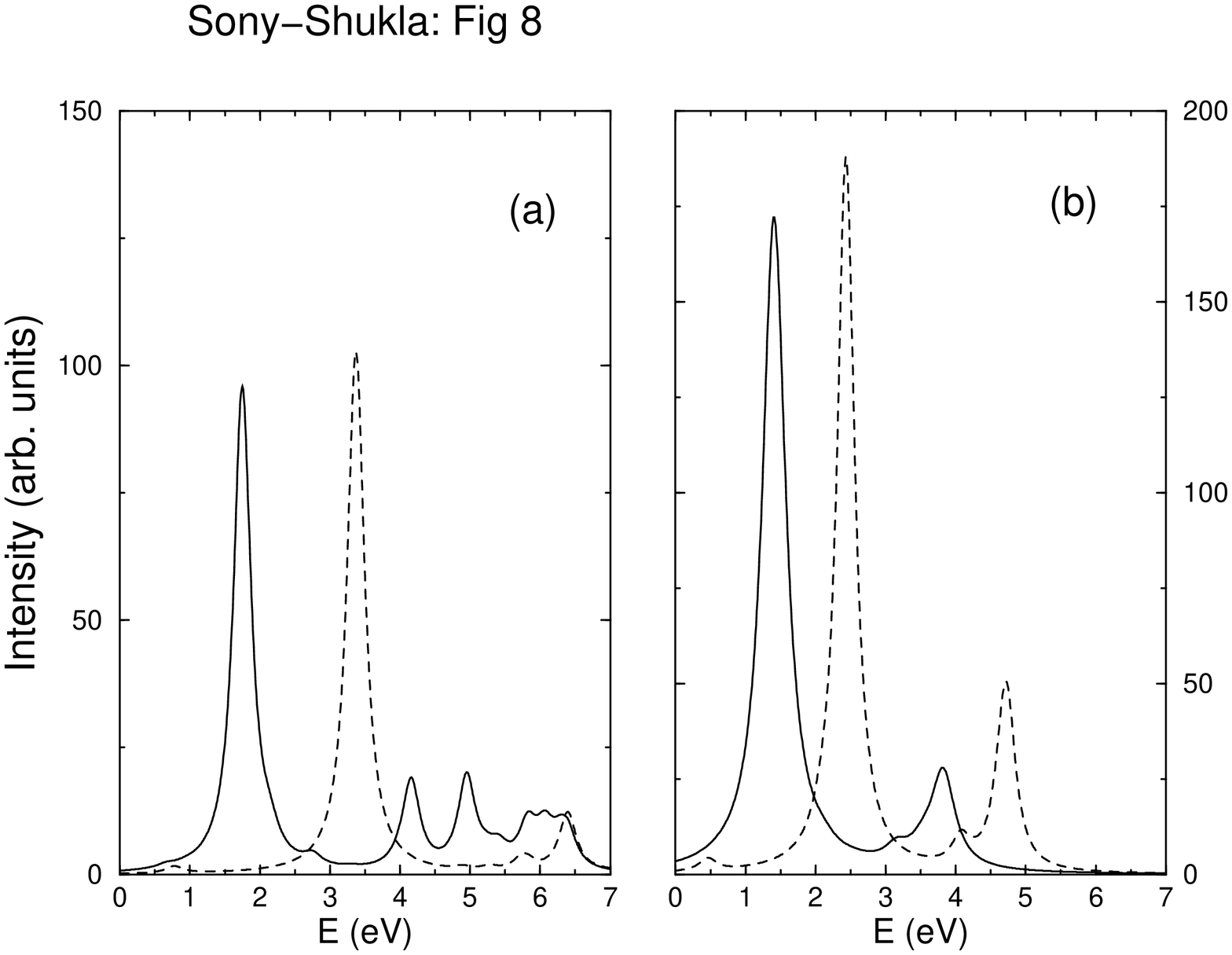}
\end{figure}

Next we compare the $2A_{g}$-PA spectra of oligo-PDPA's with polyenes
as presented in Fig. \ref{tpa-pa-2ag}. The spectra presented were
computed using the QCI approach, and the standard parameters in the
PPP Hamiltonian, for all the oligomers. The peak positions in the
oligo-PDPA spectra in this figure are blue shifted as compared those
computed with the screened parameters. For example, the locations
of the first peaks in the PDPA-5/PDPA-10 spectra are at 1.74/1.41
eV, as compared to their screened parameter locations of 1.35/1.00
eV. This suggests that the same trends will also hold for more sophisticated
MRSDCI calculations, thus taking them outside the range of the experimental
location of PA$_{g}$. As far as comparison of the $2A_{g}$-PA spectra
of oligo-PDPA's and polyenes is concerned, the shapes of the spectra
of the two materials look qualitatively similar, although quantitatively
speaking PDPA spectra are extremely redshifted as compared to their
polyene counterparts. However, upon closer examination, we find that
the nature of the excited states contributing the first peaks in these
spectra are completely different in polyenes as compared to oligo-PDPA's.
Going by the definition of the $nB_{u}$ state as the one which leads
to a strong peak in the THG spectrum,\cite{sumit} we have identified
the first and the most intense peak of oligo-PDPA $2A_{g}$ spectrum
as the $nB_{u}$ state. However, the first and the most intense peak
of the calculated $2A_{g}$-PA spectra of polyenes is not due to the
$nB_{u}$ state, but rather due to a state which is considerably lower
in energy than the $nB_{u}$ state, and is invisible in the THG spectra.
For example, for the five double-bond polyene the first peak is due
to the $3B_{u}$ state located at 6.79 eV, while the $nB_{u}$ state
for that polyene is located at 9.25 eV, with respect to the ground
state. Similarly, for the ten double-bond polyene the first peak is
again due to its $3B_{u}$ state located at 5.50 eV, while the $nB_{u}$
state is at 7.14 eV. This peculiar aspect of the $2A_{g}$-PA spectra
of polyenes will be discussed elsewhere in detail.\cite{shuk-maz}
However, here we would like to emphasize that in the $2A_{g}$-PA
spectra of oligo-PDPA's, the $nB_{u}$ has a strong presence, while
in that of polyenes another lower energy $B_{u}$ state contributes
to the spectra with large intensity, while the $nB_{u}$ state is
essentially invisible.

\section{Summary and Conclusions}

\label{conclusion}

In conclusion, we presented correlated calculations of PA spectra
of oligo-PDPA's, and compared our theoretical results to the recent
experimental results of Korovyanko \emph{et al}.\cite{koro} Good
qualitative agreement was obtained between the calculated spectra,
and the measured ones, for oligomers. As far as the specific aspects
of the spectra are concerned, PA1 feature of the $1B_{u}$ PA spectrum,
and the PA$_{g}$ feature of the $2A_{g}$ PA spectrum have been unambiguously
identified as the $mA_{g}$ and $nB_{u}$ states of the polymer, respectively.
In both the cases, the polarization of the photons involved in the
transition have been predicted to be mainly along the conjugation
direction. Our confidence regarding these predictions stems from the
fact that various calculations performed on both PDPA-5 and PDPA-10
were all in excellent agreement with each other as far as the nature
of the many-particle wave functions of these states, and the polarizations
of the photons are concerned. The only peak regarding which our predictions
at present are somewhat uncertain is the PA2 feature of the $1B_{u}$
PA spectrum. Our MRSDCI calculations on PDPA-5, identified it with
excited states whose wave functions consist of significant contributions
from the chain-based orbitals close to the Fermi level to the high-lying
phenylene-based $d/d^{*}$ orbitals.

Regarding the quantitative aspects, we summarize the results of all
our calculations, and their comparison to experiments, in table \ref{tab-sum}.
From the table it is clear that overall, the results of MRSDCI calculations
performed on PDPA-5 are in very good quantitative agreement with the
experimental results as far as the peak positions are concerned. Combined
with the assertion by the experimentalists that most of the oligomers
are five to seven repeat units long,\cite{fujii2,gontia2} one is
tempted to believe that the present theory describes the experimental
situation quite well. But, we believe that such speculations are immature
until the time the experimental results are also available on the
polarizations of the photons involved in the transitions leading to
various peaks. Therefore, it will be extremely helpful if future experiments
are performed on oriented samples so that the polarizations of the
photons involved in the transition can be measured, and the theory
presented here is tested more stringently.

\begin{table}

\caption{Summary of theoretical results, and their comparison to the experimental
values,\cite{koro} where possible. For polarization directions, $x$
implies along the chain, while $y$ implies perpendicular to it. For
calculations where more than one states were possible candidates for
an experimental feature (see text), information is provided for all
of them. Mean conjugation length of the oligomers in experiments is
believed to be five to seven repeat units.\label{tab-sum}}

\begin{tabular}{|c|c|c|c|c|}
\hline 
Oligomer&
Method&
Peak Position (Theory)&
Polarization (Theory)&
Peak position (Exp.)\tabularnewline
\hline
\hline 
PDPA-5&
QCI&
1.38 eV ($mA_{g}$)&
mainly $x$&
1.1--1.2 eV (PA1)\tabularnewline
PDPA-5&
MRSDCI&
1.19 eV ($mA_{g}$)&
mainly $x$&
\tabularnewline
PDPA-10&
QCI&
0.81--1.12 eV ($mA_{g}$)&
mainly $x$&
\tabularnewline
\hline 
PDPA-5&
QCI&
2.83 eV&
mixed&
2.0 eV (PA2)\tabularnewline
PDPA-5&
MRSDCI&
1.72--2.17 eV&
mixed(larger $y$)/mainly $x$&
\tabularnewline
PDPA-10&
QCI&
1.99 eV&
mixed&
\tabularnewline
\hline
PDPA-5&
QCI&
1.35 eV ($nB_{u}$)&
mainly $x$&
1.7 eV (PA$_{g}$)\tabularnewline
PDPA-5&
MRSDCI&
1.59 eV ($nB_{u}$)&
mainly $x$&
\tabularnewline
PDPA-10&
QCI&
1.00 eV ($nB_{u}$)&
mainly $x$&
\tabularnewline
\hline
\end{tabular}
\end{table}

\begin{acknowledgments}
This work was supported by grant no. SP/S2/M-10/2000 from Department
of Science and Technology (DST), Government of India.
\end{acknowledgments}


\begin{thebibliography}{10}
\bibitem{notation}The many-particle wave functions of half-filled conjugated polymers
such as PDPA exhibit particle-hole (charge conjugation) symmetry when
treated using the models such as the Hubbard model and the PPP model.
Therefore, for every configuration which contributes to its wave function,
its charge-conjugated (c.c.) counterpart will contribute in equal
measure. For example, if configuration $|H\rightarrow L+1\rangle$
occurs in the wave function with coefficient 0.69, configuration $|H-1\rightarrow L\rangle$
will also occur in the wave function with coefficient of the same
magnitude (although, the sign could be reversed). Thus, we will use
shorthand notation $|H\rightarrow L+1\rangle+c.c.$ (0.69) to denote
the contribution of these configurations to the wave function. Above,
$H/L$ refer to HOMO/LUMO orbitals.
\end{thebibliography}
\end{document}